\documentclass[journal=mamobx,manuscript=article]{achemso}

\title[Exact Partition Function Zeros]
{Exact Partition Function Zeros of a Polymer on a Simple-Cubic Lattice}

\author{Jae Hwan Lee}
\affiliation[Soongsil University]
{School of Systems Biomedical Science and Department of Bioinformatics and Life Science, Soongsil University, Seoul 156-743, Korea}

\author{Seung-Yeon Kim}
\email{sykimm@ut.ac.kr}
\affiliation[Korea National University of Transportation]
{School of Liberal Arts and Sciences, Korea National University of Transportation, Chungju 380-702, Korea}

\author{Julian Lee}
\email{jul@ssu.ac.kr}
\affiliation[Soongsil University]
{School of Systems Biomedical Science and Department of Bioinformatics and Life Science, Soongsil University, Seoul 156-743, Korea}

\begin{document}

\begin{abstract}
We study conformational transitions of a polymer on a simple-cubic lattice by calculating the zeros of the exact partition function, up to chain length 24. In the complex temperature plane, two loci of the partition function zeros are found for longer chains, suggesting the existence of both the coil-globule collapse transition and the melting-freezing transition. The  locus corresponding to coil-globule transition clearly approaches the real axis as the chain length increases, and the transition temperature could be estimated by finite-size scaling. The form of the logarithmic correction to the scaling of the partition function zeros could also be obtained. The other locus does not show clear scaling behavior, but a supplementary analysis of the specific heat reveals a first-order-like pseudo-transition.
\end{abstract}

\maketitle

%%%%%
\section{Introduction}
%%%%%

Conformational transitions of a polymer have been a subject of interest for many years~\cite{Li78,Kho81,dG75,dG78,dG79,D82,MMG73,FJB75,JB76,R74,R77,R85,M77,CS80,B84,ML90,SOP92,GH95,YLH96,TROW96,G97,FG97,PWF98,TL98,YP00,L04,LKL10,LKL11a,Zhou96,RPB05,PW06,VBJ07}. The coil-globule transition of a homopolymer, also called the collapse transition, has been most widely studied. A polymer chain in a dilute solution is influenced by both hydrophobic interactions between the monomers and the excluded volume effect, which cancel each other at a special temperature  $T=T_\theta$. The polymer chain adopts expanded and compact conformations for $T > T_\theta$ and $T < T_\theta$ respectively, with $T= T_\theta$ being the temperature of coil-globule collapse transition. $T > T_\theta$ region of the polymer system can be mapped into a magnetic system where the number of components is formally set to zero, with the infinite chain length corresponding to the critical temperature\footnote{The temperature of the magnetic system is conjugate of the polymer chain length $N$, and it has nothing to do with the temperature $T$ of the polymer model.}~\cite{dG75,dG79}. Then $T= T_\theta$ is a tricritical point where both of the quadratic and the quartic terms in the corresponding field theory vanish~\cite{dG79}.   

The radius of gyration (or the end-to-end distance) $R$ of a polymer chain with $N$ monomers, near $T=T_\theta$, 
is generally expressed by the scaling theory~\cite{dG79}, 
\begin{equation}
	\langle R^2 \rangle \sim N^{2\nu_t} f(\tau N^{\phi}),
\label{scaling}
\end{equation}
where the exponent $\nu_t$ represents the geometrical properties of a polymer at the tricritical point, 
and the exponent $\phi$, called the \textit{crossover} exponent~\cite{Grif73,Cardy96}, 
measures how rapidly the system undergoes the transition as the temperature $T$ approaches the tricritical temperature $T_\theta$.
The reduced temperature is defined as $\tau \equiv \left \vert T-T_\theta \right \vert / T_\theta$, scaling as $\tau \sim N^{-\phi}$ as $T \to T_\theta$.
The scaling function $f(x)$ behaves as follows~\cite{ML90}
\begin{eqnarray}
	f(x) &=& \left\{
	\begin{array}{lll}
		x^{(6/(d+2)-2\nu_t)/\phi}	& \quad &	\mathrm{if}~~x \to \infty,	\\
		\mathrm{const.}				& &			\mathrm{if}~~x \to 0,		\\
		|x|^{(2/d - 2\nu_t)/\phi}		& &			\mathrm{if}~~x \to -\infty.

	\end{array}
		\right. \label{leading}
\end{eqnarray}
Since the upper tricritical dimension is three~\cite{PB89}, the tricritical exponent can be obtained from the mean field theory as $\phi=1/2$ and $\nu_t=1/2$, but the scaling form (\ref{scaling}) is expected to have logarithmic corrections~\cite{dG78}.

In addition to the coil-globule collapse transition which is of second order in three dimensions~\cite{M77}, a first-order transition from liquid-like globule to a solid-like phase is expected at a temperature lower than $T_\theta$~\cite{Zhou96,RPB05,PW06,VBJ07}, which has been the subject of active research recently but less well understood than the coil-globule collapse transition.

In this work, we enumerate the number of all possible conformations on a simple-cubic lattice up to chain length 24. In particular, we study the zeros of the exact partition function, which are much more sensitive indicators of phase transition than the real-valued quantities such as specific heat. We observe that for chain length longer than 16 there are two distinct loci of zeros, suggesting the existence of both the coil-globule collapse transition and liquid-solid melting-freezing transition. The exact partition function zeros of three-dimensional lattice polymer have been studied for simple~\cite{FJB75} and face-centered~\cite{R77} cubic lattices for chain lengths up to 13 and 10, respectively, but these lengths  were too short to reveal two distinct loci, and the results for longer chains up to 31~\cite{FJB75} were based on Monte Carlo sampling with limited accuracy.

Although the chains whose conformations are exhaustively enumerated in this work are still much shorter than those studied with Monte Carlo samplings, the exactness of the calculation enables us to extrapolate the finite $N$ data to obtain the result in the limit of infinite chain length with a reasonable accuracy.  The chains were still too short to extract the large $N$ behavior of the inner locus corresponding to the melting-freezing transition, but the tricritical collapse temperature $T_\theta$ could be obtained from the scaling behavior of the outer locus. The form of the logarithmic correction to the scaling of the partition function zeros could also be obtained. Additional analysis was performed on the exact specific heat, which shows the first-order-like pseudo-transition near the inner locus, in accordance with the previous result based on a chain-growth sampling~\cite{VBJ07}.

%%%%%
\section{The Number of Conformations}
%%%%%

Conformations of a polymer chain with $N$ monomers are modelled by  self-avoiding walks with $N-1$ steps on a simple-cubic lattice.
The position of a monomer $i$ is expressed as ${\bf r}_i = (a,b,c)$ with integer values of the coordinates $a$, $b$, and $c$. The coordinates satisfy the constraints  $\vert {\bf r}_i-{\bf r}_{i+1} \vert = 1$ and ${\bf r}_i \neq {\bf r}_j$ for $i \neq j$ due to the chain connectivity and the excluded volume effect. 

We consider the Hamiltonian with the nearest-neighbor interaction:
\begin{equation}
    {\cal H} = -\epsilon \sum_{i<j} \Delta ({\bf r}_i, {\bf r}_j), \label{modelH}
\end{equation}
where
\begin{equation}
\Delta ({\bf r}_i, {\bf r}_j) = \left\{
	\begin{array}{ll}
		1 & ~~{\rm if} ~~ |i-j| > 1 ~~ {\rm and} ~~ |{\bf r}_i-{\bf r}_j| =1,\\
		0 & ~~{\rm otherwise},
	\end{array}	\right.
\end{equation}
and $\epsilon$ is set to a positive value to incorporate the attractive interaction between the monomers.

Assuming that the polymer chain has an intrinsic direction, 
the conformations with reverse labels $i \leftrightarrow N-i+1$ 
for all $(i=1,2, \cdots, N)$ are considered distinct. 
For a generic conformation, the rigid rotations and reflections
form an 48-fold symmetries in three dimensions.
Exceptions are the cases of lower-dimensional conformations embedded in higher dimensional spaces. In three dimensions, 24-fold and 6-fold symmetries exist for the planar and linear conformations, since they are invariant under transformation perpendicular to the underlying plane and straight line. Therefore reduced numbers of conformations, where conformations related by  symmetry are counted only once,  are computed in order to prevent the waste of computational resources~\cite{LKL11b}. Since the energy (\ref{modelH}) depends only on the number of inter-monomer contacts $K$, we classify the conformations according to the value of $K$. The number of conformations  $\Omega^{(d)}(K)$ for a given contact $K$ in $d$ dimensions, with discrete rotations and reflections considered distinct, is obtained from the reduced number of conformations $\omega^{(d)}(K)$ as
~\cite{LKL11b}
\begin{equation}
\begin{array}{l}
	\Omega^{(1)} (K) = 2  \omega^{(1)} (K) = 2 \delta_{K,0}, \\
	\Omega^{(2)} (K) = 8  \omega^{(2)}  (K)+ 4  \omega^{(1)}(K), \\
	\Omega^{(3)} (K) = 48  \omega^{(3)} + 24  \omega^{(2)}  (K)+ 6  \omega^{(1)}(K).
\end{array}
\label{total}
\end{equation}
From here on, we drop the dimension index and use $\Omega(K)$ to denote $\Omega^{(3)}(K)$. The number of conformations $\Omega(K)$ for $N=24$ is presented 
in Table~\ref{table:one}.

%%%%%
\section{Partition Function Zeros in the Complex Temperature Plane}
%%%%%

Partition function zeros have been the subject of interest as a sensitive indicator of a phase transition~\cite{L04,LKL10,LKL11a,YL52,F65,IPZ83,AH00,WW03,CL05,BDL05}. Partition function zeros were introduced by Yang and Lee in the complex fugacity plane of a fluid system and the complex magnetic-field plane of the nearest-neighbor Ising ferromagnet (Yang-Lee zeros), to study the phase transition driven by the fugacity or the magnetic field~\cite{YL52}.
Later, Fisher~\cite{F65} used the partition function zeros in the complex temperature plane 
(Fisher zeros)
of the square-lattice Ising model, to study the temperature driven transition. In the thermodynamic limit, the locus of zeros forms a continuous curve which crosses the real axis if a transition exists. Thus, the theory of partition function zeros provides the explanation on how the partition function, which is an analytic function of thermodynamic parameters at a finite size, acquires the singularities necessary for a phase transition in the thermodynamic limit. In the case of Fisher zeros, the transition temperature in the thermodynamic limit is  the intersection point of the locus of zeros with the real temperature axis. Therefore, the conjugate pair of zeros closest to the positive real axis, called the \textit{first zeros}, determine the leading singular behavior of the partition function. Since the behavior of the first zeros can be analyzed separately from the other zeros, the phase transition can be analyzed more accurately by computing the partition function zeros than studying real-valued quantities such as the specific heat which includes the effect from all the zeros.

In the case of the lattice polymer of our interest, the partition function is written in terms of $\Omega(K)$ as  
\begin{equation}
    Z(y) =\sum_{\{\sigma\}} e^{-\beta \cal H} = \sum_{K} \Omega(K) y^{-K}, \label{polPF}
\end{equation}
where $\{\sigma\}$ denotes a sum over all possible conformations, $\beta = 1/k_B T$, and $y \equiv \exp(-\beta \epsilon)$ which ranges between 0 and 1 for positive $T$;
$y=0$ when $T=0$ and $y=1$ when $T \to \infty$. 
Since the maximum number of contacts  $K_M$ is finite for a given chain length $N$, $y^{K_M} Z(y)$ is a polynomial of order  $K_M$ in $y$. Therefore the partition function zeros in the complex $y$ plane can be obtained by solving the  polynomial equation $y^{K_M} Z(y)=0$ with \textsc{mathematica}.
 
Figure~\ref{zeros} shows the partition function zeros 
in the complex temperature plane for several values of chain lengths. 
The result for $N = 13$ (Fig.~\ref{zeros}(a)) agrees with the previous result obtained by Finsy \textit{et al.}~\cite{FJB75} by exact enumeration after a change of variable, but the chain length is too short to reveal two distinct loci of zeros. The partition function zeros for $14 \le N \le 31$ were studied in the same reference by Monte Carlo sampling, but again two distinct loci were not visible. One of the reasons is that the zeros in the complex temperature plane of $z=e^{\beta \epsilon}$ were investigated there instead of $y=e^{-\beta \epsilon}=1/z$ as in the current work. In the former case, the temperature range of interest $T>0$ corresponds to the real line with $z>1$, with $T=0$ corresponding to $z=+\infty$. Therefore, features relevant to low temperature behavior of the system may be easily missed if only a finite region near origin is considered. In contrast,  $T>0$ corresponds to the segment $0 < y < 1$ in the $y$ plane, and consequently all of the physical regions can be examined with ease.

%Even in figure 2 of ref.\cite{FJB75}, a small outward protrusion in the locus of is observed, which may eventually meet the real axis in the very large positive value of $z$ which is far out of the range of the figure. 
 The exact zeros plotted in the $y$ plane clearly exhibit the splitting of the locus into two distinct branches already for $N \ge 16$ (Fig.~\ref{zeros}(b) and (c)). Since there are two visually distinct loci of zeros, we will select the first zeros for each locus and analyze them separately in the following sections.

%%%%%%%%%%
\section{The Outer Locus and the Coil-Globule Collapse Transition}
%%%%%%%%%%

The first zeros of the outer locus approach the real axis suggesting a non-trivial transition in the thermodynamic limit (Fig.~\ref{nre}(a) (open circles)). From the scaling of the partition function whose form is similar to (\ref{scaling}), the first zeros $y_1$ are expected to scale as~\cite{LKL10,IPZ83}:
\begin{equation}
    y_1(N)-y_\theta \sim N^{-\phi},
\label{ycfull}
\end{equation}
where $y_\theta \equiv \exp(-\epsilon/k_B T_\theta)$. In three dimensions corresponding to the upper tricritical dimension, the scaling form (\ref{scaling}) is modified by logarithmic corrections~\cite{dG78,D82}, and consequently (\ref{ycfull}) is expected to be modified to be of the form
\begin{equation}
    y_1(N)-y_\theta \sim N^{-\phi} (\log N)^{-\lambda}.
\label{yclog}
\end{equation}
The power of logarithmic correction $\lambda$ can be estimated from the imaginary part of (\ref{yclog}):
\begin{equation}
    {\rm Im} [y_1(N)] \sim N^{-\phi} (\log N)^{-\lambda},
\label{ycim}
\end{equation}
which is rewritten as
\begin{equation}
    \mathrm{Im} [y_1(N)] N^{\phi} \sim (\log N)^{-\lambda}.
\label{ycim2}
\end{equation}
Taking the logarithm, we get
\begin{equation}
   \log ({\rm Im} [y_1(N)] N^{\phi}) \simeq - \lambda \log \log N + {\rm constant}.
\label{ycim3}
\end{equation}
In order to remove the constant term, the difference of (\ref{ycim3}) for neighboring chain lengths is taken:
\begin{equation}
   \log \left(\frac{{\rm Im} [y_1(N+2)] (N+2)^{\phi}}{{\rm Im} [y_1(N)] N^{\phi}}\right) \simeq -\lambda \log \left( {\frac{\log (N+2)}{\log N}} \right),
\label{ycim4}
\end{equation}
where we used $N+2$ instead of $N+1$ because even and odd number of chain lengths are expected to exhibit distinct scaling behaviors~\cite{LKL10}. 
From (\ref{ycim4}) the finite-size estimation of $\lambda$ is obtained as:
\begin{equation}
	\lambda(N) \simeq  -\frac{\log({\rm Im} [y_1(N+2)] (N+2)^{\phi})/ {\rm Im} [y_1(N)] N^{\phi})}{ \log ( \log(N+2) / \log N))}  \label{ycimlog}
\end{equation} 
with $\phi=1/2$.  
We applied Bulirsch-Stoer (BST) method~\cite{BST} for the data of even $N$ with $12 \le N \le 24$ to estimate the value in the limit $N \to \infty$. We obtained
\begin{equation}
	\lambda = 0.642(28), \label{lamest}
\end{equation}
where the error is estimated  by examining the robustness of the extrapolated value  with respect to perturbations of the data points~\cite{LKL10}. The data points are chosen to maximize this robustness. 
The estimated value (\ref{lamest}) exhibits remarkable agreement with
\begin{equation}
	\lambda=\frac{7}{11} = 0.636346, \label{lamboyl}
\end{equation} 
appearing in the scaling of the Boyle temperature $T_B$ where the second virial coefficient vanishes~\cite{dG78,GH95,VBJ07}:
\begin{equation}
	T_B (N) - T_\theta \sim N^{-1/2} (\log N)^{-7/11}.
\end{equation} 
We then estimate the collapse temperature $T_\theta$ by taking the real parts of the scaling relations (\ref{ycfull}) or (\ref{yclog}).
Applying the BST extrapolation for the data of even $N$ with $14 \le N \le 24$, the collapse temperature is obtained as $y_\theta = 0.7185(94)~(k_B T_\theta / \epsilon = 3.03(12))$ without the logarithmic correction, and  $y_\theta = 0.7653(174)~(k_B T_\theta / \epsilon = 3.76(32))$ with the correction factor of $(\log N)^{-7/11}$(Fig.~\ref{nre}(b)). As shown in Table~\ref{table:two}, 
the agreement with previous results is better when the logarithmic correction is included. It is to be noted that although the  maximum length of chain 24 considered here is much less than those in Monte Carlo studies which is as long as 32000~\cite{VBJ07}, the exactness of our data enables us to perform extrapolation to large $N$ with reasonable accuracy. 

We also note that the intersection of the locus of zeros with the real axis is estimated to be at a temperature higher than $k_B T_\theta / \epsilon = 1.81(2)$ obtained by fitting the Monte Carlo data of partition function zeros to a polynomial curve~\cite{FJB75}. It was conjectured in the same reference that the corresponding transition is of first-order, which was later conjectured to be a low temperature melting-freezing transition~\cite{M77}. However, we found no evidence that the transition is of the first order. Since there is a separate inner locus corresponding to a transition at a lower temperature, and the intersection of the outer locus with the real axis in the thermodynamic limit is now estimated to be at 3.76 rather than 1.81, the outer locus seems to be the one corresponding to the coil-globule collapse transition, which is well established to be a second-order transition~\cite{Li78,Kho81,M77}.

%%%%%
\section{The Inner Locus and the Melting-Freezing Transition}
%%%%%

The scaling of inner locus is much worse than that of the outer locus in this range of chain lengths. The distribution of the first zeros for various chain lengths is rather irregular (Fig.~\ref{nre}(a)), and the real part shows oscillatory behavior as a function of $N^{-1/2}$ (Fig.~\ref{nre}(b)). Therefore no reasonable estimate of the transition temperature in the thermodynamics limit could be made. This irregular behavior of the melting-freezing transition point is also in agreement with the previous result obtained by computing the specific heat with chain-growth sampling for $N$ up to 125~\cite{VBJ07}. This irregularity is understandable, as explained in the same reference. The effect of melting-freezing transition which is expected to survive in the limit of large $N$ limit, and that of the excitation pseudo-transition which appears only near a special value of $N$ called magic numbers, are intermixed to give a rather complex behavior. 

To complement the result from the partition function zeros, we performed additional analyses by computing the exact specific heat per monomer, 
\begin{equation}
	\frac{C_V(T,N)}{N k_B} = \frac{1}{N k_B}\frac{\partial E}{\partial T}
	= \frac{\beta^2}{N} \frac{\partial^2 \ln Z}{\partial \beta^2}
	= \frac{(\ln y)^2}{N}\left( \langle K^2 \rangle - \langle K \rangle^2  \right),
\end{equation}
which is plotted in Fig.~\ref{specific}(a) as a function of $y$ for several values of $N$. The effect of the melting-freezing and excitation transitions are mixed to manifest themselves as one prominent peak. In accordance with the previous result from a chain growth sampling~\cite{VBJ07}, we find that due to the effect of the excitation transition at the magic numbers $N_c=8,12,18, \cdots$ satisfying $N_c = L^3$ or $N_c=L^2(L \pm 1)$, the peak becomes sharper as $N$ is increased from $N_c-1$ to $N_c$  and becomes flat again at $N=N_c+1$ (Fig.~\ref{specific}(a)).  As to be expected, the positions and the values of the peak exhibit irregular behavior as functions of the chain length (Fig.~\ref{specific}(b)), similar to the result from the chain growth sampling~\cite{VBJ07}. The peak positions are in the range $0.10 < y_{\rm peak} < 0.25$,  much lower than the coil-globule collapse transition temperature $y_c = 0.7653(174)$ obtained from the analysis of the outer locus of the partition function zeros, but much closer to the inner locus of zeros, whose real values are distributed mostly in the range $0.0 \le y < 0.1$ (Fig.~\ref{nre}(a)). 

\begin{figure}
\includegraphics[width=\textwidth]{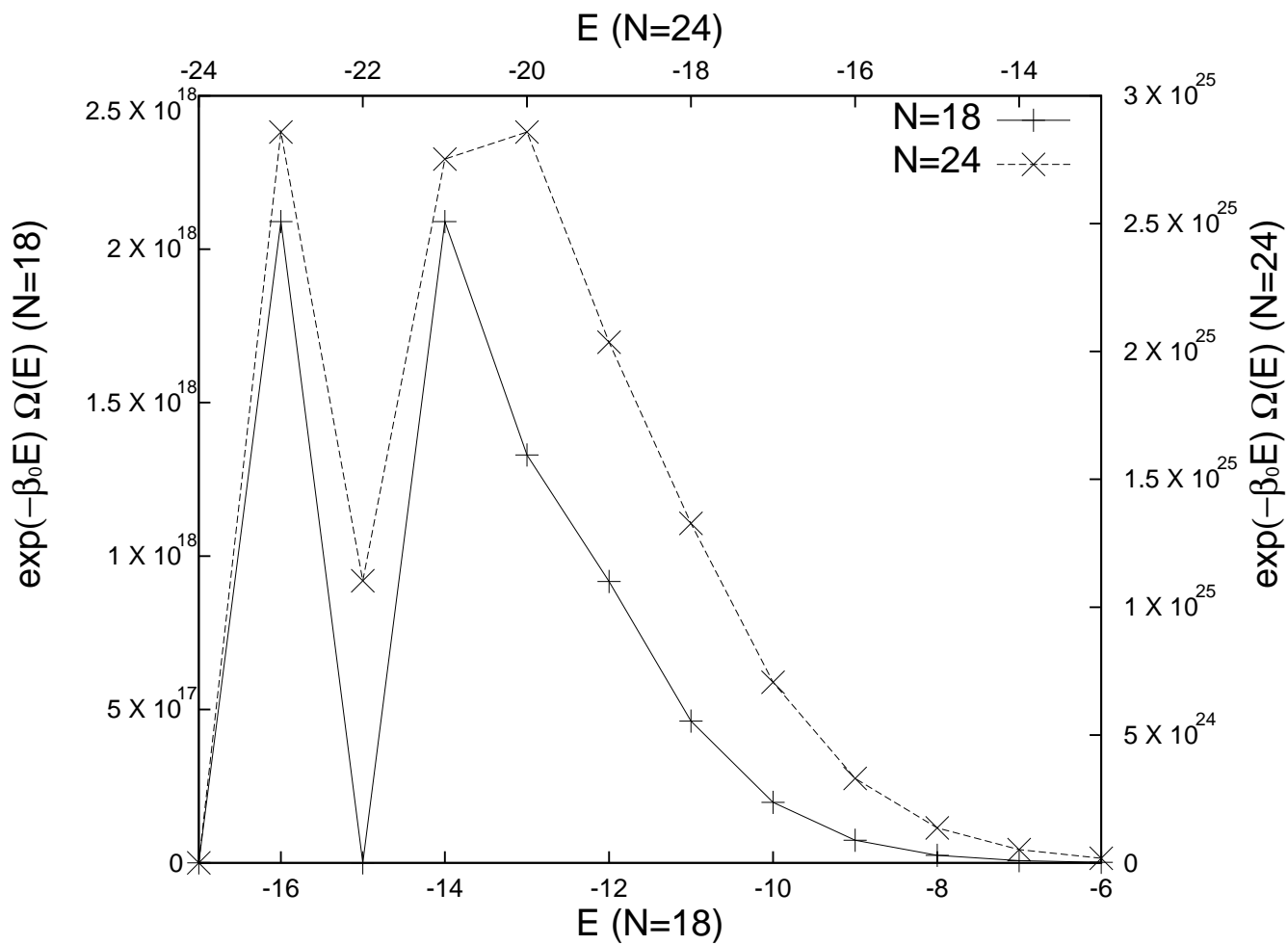}
\caption{The density of states multiplied by Boltzmann factor $e^{-\beta_0 E} \Omega(E)$ as a function of $E$, with $\beta_0$ near the transition temperature, for $N=18$ and $24$. The values of $\beta_0$ are adjusted for each $N$ to make the heights of the two peaks equal. The double peaks show a clear sign of first-order-like transition.}
\label{peaksfig}
\end{figure}

In the microcanonical formalism, the first-order phase transition is signaled by the existence of a region of energy with $\frac{\partial^2 S}{\partial E^2} > 0$ where $S(E)=\log \Omega(E)$\footnote{This corresponds to a negative value of microcanonical specific heat $C_\mathrm{micro} \equiv - \beta^2 \left(\frac{\partial^2 S(E)}{\partial E^2}\right)^{-1}<0$.}, 
which is invisible in the canonical formalism~\cite{Gross01,Jung06}. For the system with discrete values of energy, this condition translates into the condition that there exists $i$ with
\begin{equation}
	S(E_{i+1})-2S(E_i)+S(E_{i-1}) > 0,
\label{dip}
\end{equation}
where $i$ labels the energy values in the ascending order.
These conditions are equivalent to the existence of $\beta_0$ near the transition temperature such that
\begin{equation}
	e^{-\beta_0 E} \Omega(E) \label{peaks}
\end{equation}
has at least two peaks corresponding to distinct phases. We see that the chains with  $N \ge 16$ as well as even value of $N \ge 8$ satisfy these properties. We plot the two peaks of $e^{-\beta_0 E} \Omega(E)$ in Fig.~\ref{peaksfig} for the largest magic number ($N=18$) and the largest number  ($N=24$) among the chain lengths we studied, where the values of $\beta_0$ were adjusted for each chain so that the heights of the peaks are the same. These features clearly shows the first-order-like nature of the pseudo-transition, and are also consistent with similar results from the chain-growth sampling~\cite{VBJ07}.

In contrast to melting-freezing and excitation transition, the effect of the coil-globule collapse transition is not readily visible in the specific heat (Fig.~\ref{specific}). This is due to the fact that the collapse transition is of the second order, where the  specific heat is continuous and only its derivative is discontinuous or divergent.

%%%%%%%%%%%%%%
\section{Discussions}
%%%%%%%%%%%%%%

We studied conformational transitions of a polymer by exhaustively enumerating the number of all possible self-avoiding walks on a simple-cubic lattice up to chain length 24. 
 Although the lengths of the chain are much smaller than those studied by Monte Carlo samplings, the strength of our result is that they are obtained from exact enumeration and hence contain no errors, enabling us to make extrapolation to infinite chain length with a reasonable accuracy. 

 Furthermore, by studying partition function zeros in the complex temperature plane, we could obtain information which is not readily available in the real-valued quantities such as specific heat. We observed two distinct loci of partition function zeros in the complex temperature plane, suggesting the existence of coil-globule collapse transition and liquid-solid melting-freezing transition. 

From the finite-size scaling of the first zeros of the outer locus with mean field crossover exponent $\phi=1/2$, the scaling form of the first zero with logarithmic correction factor was conjectured to be $y - y_\theta \sim N^{-1/2} (\log N)^{-7/11}$ as in the case of the scaling of the Boyle temperature. The  collapse transition temperature was estimated to be $k_B T_\theta / \epsilon = 3.03(12)$ and  $k_B T_\theta / \epsilon = 3.76(32)$ with and without logarithmic correction, respectively. The result shows better agreement with previous results in the presence of the logarithmic correction, suggesting that the conjectured form of the logarithmic correction to the scaling of the first zeros is indeed correct. 

\begin{figure}
\includegraphics[width=\textwidth]{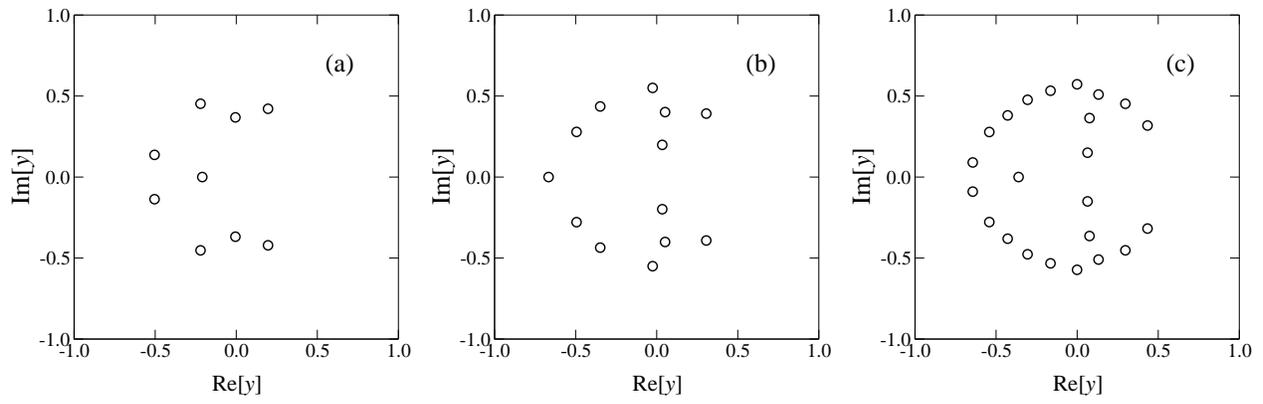}
\caption{Distribution of the partition function zeros 
in the complex temperature ($y = e^{-\beta \epsilon}$) plane
for (a) $N=13$, (b) $N=16$, and (c) $N=24$.
}
\label{zeros}
\end{figure}

\begin{figure}
\includegraphics[width=\textwidth]{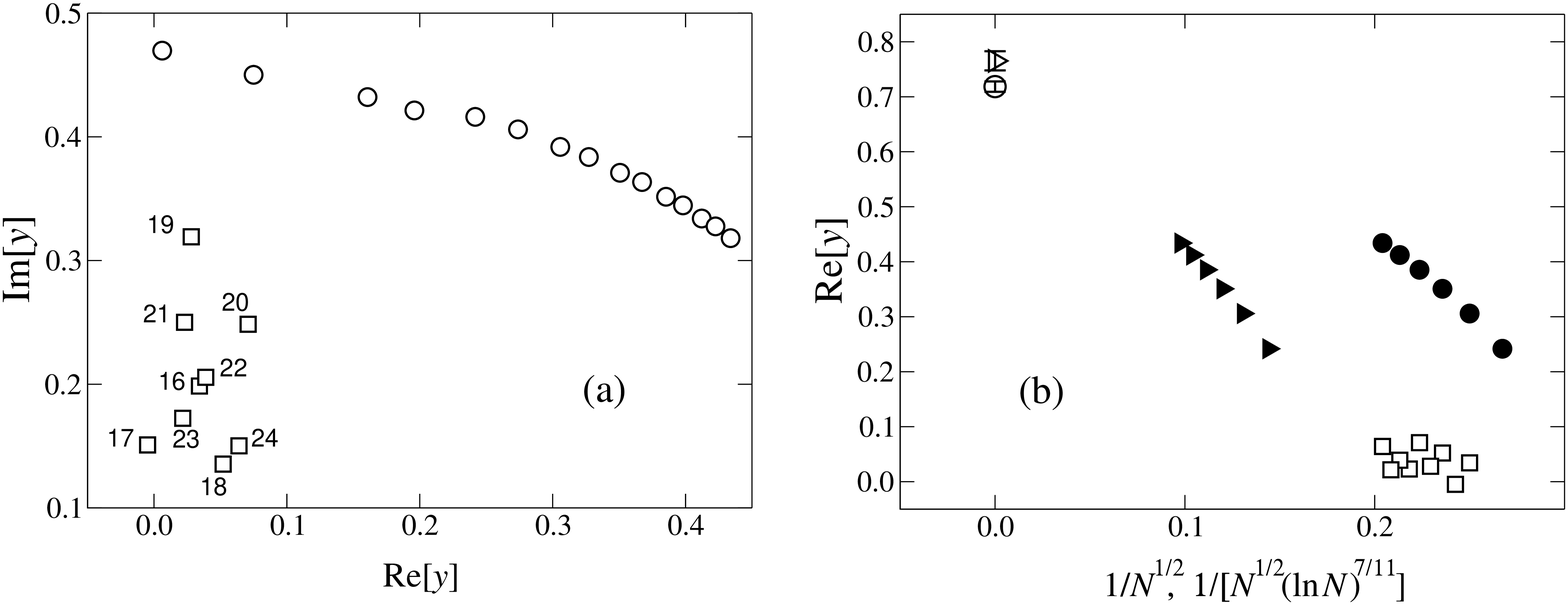}
\caption{(a) The first zeros of outer locus in Fig.~\ref{zeros} for $N=10,~11,~12,\cdots, ~24$ are plotted as open circles, 
and those of inner locus for $N=16,~17,~18,\cdots, ~24$ as open squares. Since the distribution of zeros is symmetric with respect to the real axis, only the first quadrant of the complex plane is shown. 
(b) Values of the real part of the first zeros of the outer locus are shown as a function of $N^{-1/2}$ (solid circles) and $N^{-1/2} (\ln N)^{-7/11}$ (solid triangles) for even $N$ with $14 \le N \le 24$. From the extrapolation to $N \to \infty\ (1/N=0$), the  collapse temperature is obtained as $y_\theta = 0.7185(94)$ without a logarithmic correction (the open circle with an error bar), and $y_\theta = 0.7653(173)$ with the correction factor of  $N^{-1/2} (\ln N)^{-7/11}$ (the open triangle with an error bar). On the other hand, those of the inner locus for $N=16,~17,~18,\cdots, ~24$ (open squares, plotted as the function of $N^{-1/2}$) show irregular behavior and cannot be extrapolated.
}
\label{nre}
\end{figure}

\begin{figure}
\includegraphics[width=\textwidth]{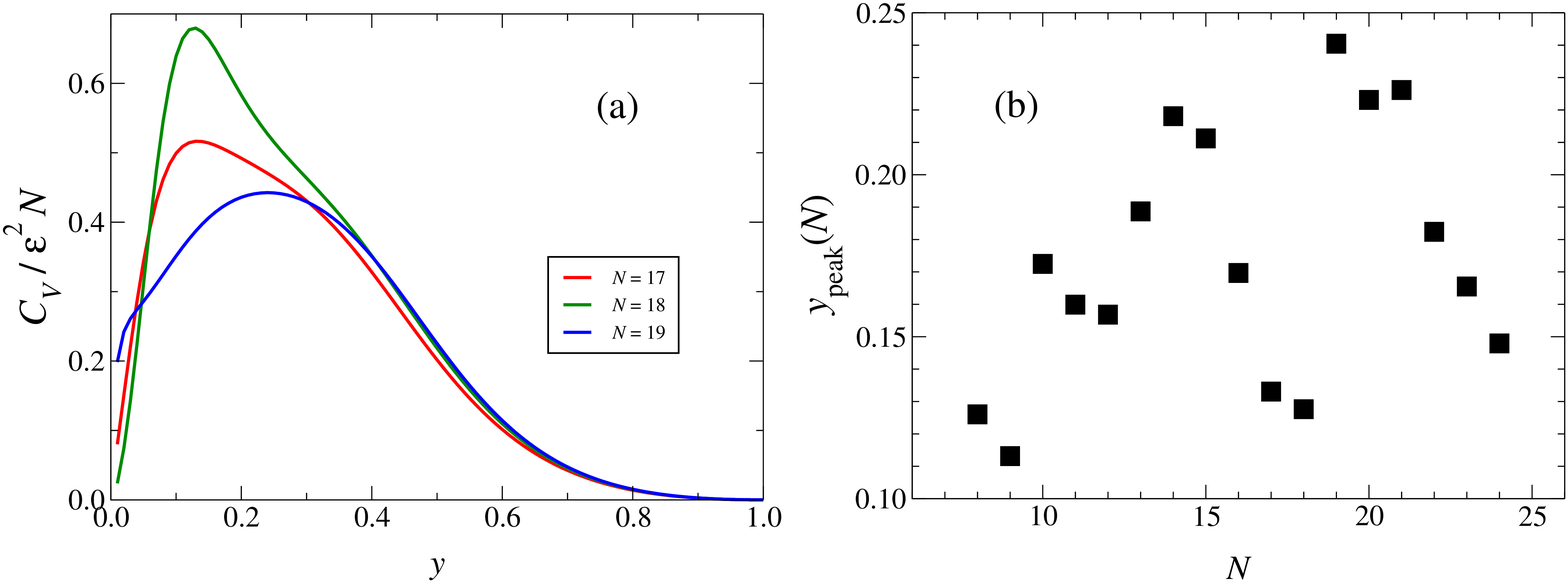}
\caption{(a) Curves of the specific heat per monomer as a function of $y$, for $N$ near the magic number $18=2 \cdot 3^2$. The peak rises sharply as $N$ is increased from 17 to 18, but becomes flat again as it is increased to 19, due to effect of the excitation pseudo-transition at $N=18$ 
(b) Peak temperatures of the specific heat for $8 \le N \le 24$ as a function of chain length $N$, 
showing an irregular pattern within $0.10 < y_{\rm peak} < 0.25$.}
\label{specific}
\end{figure}

The results for the collapse transition indicate an additional advantage  of studying the partition function zeros. Being a second-order transition where the specific heat is finite and continuous, the signal for the transition cannot be easily detected by examining the peak of the specific heat (Fig.~\ref{specific}(a)). On the other hand, since the partition function zeros is due to the singularities of the partition function regardless of the order of the transition, the signal for the transition is clearly visible as a locus of zeros approaching the real axis as the chain length increases (Fig.~\ref{zeros} and \ref{nre}), making them an indispensable tool for studying phase transitions.

The behavior of the inner locus was not regular enough for the current chain lengths to estimate the melting-freezing temperature in the limit of infinite chain lengths, due to mixture of the effect from the finite-size excitation pseudo-transition. The existence of the energy region with the negative value of microcanonical specific heat, which manifests itself as double peaks in the density of states multiplied by Boltzmann factor near transition temperature, shows a clear sign of first-order-like nature of the pseudo-transition.

The excitation transition is analogous to the folding transition of HP protein, which appears only for a particular sequence~\cite{CKI99,BJ03,BJ05}. It would be interesting to compare the transition behaviors of designable proteins, random heteropolymers and homopolymers, all on the same lattice.

It is to be  noted that although partition function zeros can be used to investigate the existence and the property of the phase transition, the information on the nature of the phases themselves cannot be obtained. Various geometric parameters of conformations contributing the phases of interest should be analyzed in order to confirm that the transitions we observe are indeed the coil-globule collapse and solid-liquid melting-freezing transitions.

%%%%%%%%%%%%%%%%
\acknowledgement
Jae Hwan Lee and Julian Lee were supported 
by Mid-career Researcher Program 
through the National Research Foundation of Korea (NRF) 
funded by the Ministry of Education, Science and Technology (MEST) 
(No. 2010-0000220), 
and
Seung-Yeon Kim was supported 
by Basic Science Research Program 
through the NRF 
funded by the MEST 
(No. 2011-0014994).

%%%%%%%%%%%%

%
%
\newpage
\begin{table}
\caption{The number of conformations $\Omega(K)$ 
as a function of the number of contacts $K$ for $N=24$.}
%\begin{ruledtabular}
\begin{tabular}{ccr}
\hline
\hline

$K$	&~~~~~& $\Omega(K)$ $\qquad$\\

\hline

0	&& 238306751550942 \\
1	&& 601441550088000 \\
2	&& 856234452257592 \\
3	&& 919771036344192 \\
4	&& 821203501326936 \\
5	&& 642255091228800 \\
6	&& 455089815998760 \\
7	&& 298905402843360 \\
8	&& 184343422767744 \\
9	&& 107615281912368 \\
\hline
10	&& 59739246931968	\\
11	&& 31649589839232 \\
12	&& 16004806431576 \\
13	&& 7677745597008 \\
14	&& 3470790178464 \\
15	&& 1454509923624 \\
16	&& 559820945808 \\
17	&& 190765562640 \\
18	&& 57066241104 \\
19	&& 13933700784 \\
\hline
20	&& 3113368896 \\
21	&& 477160080 \\
22	&& 30437280 \\
23	&& 12554256 \\

\hline
Total	&&	5245988215191414 \\
\hline
\hline
\end{tabular}
%\end{ruledtabular}
\label{table:one}
\end{table}
\newpage
\begin{table}

\caption{The values of the coil-globule collapse temperature $k_B T_\theta/\epsilon$ of a lattice polymer on a simple-cubic lattice with nearest neighbor interaction, obtained in the current work (first two lines), are compared with previous results on the same model.}
%\begin{ruledtabular}
\begin{tabular}{lrc}
\hline
\hline
Method & $N_{\max}$ & $k_B  T_\theta/\epsilon$ \\

\hline
Exact PFZ (without $(\log N)^{-7/11}$ correction)& 24 & {\bf 3.03(12)} \\
Exact PFZ (with $(\log N)^{-7/11}$ correction)& 24 & {\bf 3.76(32)} \\
RCG~\cite{MMG73}	& 2000 & 3.64 $\sim$ 4.13 \\
MC PFZ~\cite{FJB75}		& 31	& 1.81(2) \\
MC~\cite{JB76}		& 40	& 3.714(11)	\\
MC~\cite{CS80}		& 299	& 4	\\
MC~\cite{B84}		& 1024	& 3.713(7) \\
SS~\cite{ML90}		& 250	& 3.65(8) \\
MC~\cite{SOP92}		& 100	& 2.972(6) \\
MC~\cite{GH95}		& 5000	& 3.721(6) \\
CBV~\cite{YLH96}	& 200	& 3.45 \\
MC~\cite{TROW96}	& 1200	& 3.598(54) \\
PERM~\cite{G97}		& 10000	& 3.724 \\
PERM~\cite{FG97}	& 2048	& 3.717(2) \\
MC~\cite{PWF98}		& 1000	& 3.71(1) \\
BGY~\cite{TL98}	& 600 &	3.745 \\
MC~\cite{YP00}		& 16000	& 3.71(1) \\
Improved PERM~\cite{VBJ07} & 32000 & 3.72(1) \\
\hline
\hline
\end{tabular}
\label{table:two}
\end{table}
%\end{tabular}
%\end{ruledtabular}
\footnotetext{PFZ: partition function zeros}
\footnotetext{RCG: Rosenbluth chain growth}
\footnotetext{MC: Monte Carlo}
\footnotetext{SS: scanning simulation}
\footnotetext{CBV: configurational-bias-vaporization}
\footnotetext{PERM: pruned-enriched Rosenbluth method}
\footnotetext{BGY: Born-Green-Yvon integral equation}

\newpage
\newpage
%
%
%
%
%\begin{figure}
%\includegraphics[width=\textwidth]{nlambda_even12.eps}
%\caption{The finite-size approximations of the exponent of the logarithmic correction $\lambda$ are shown as a function of $1/N$ for even $N$ with $12 \le N \le 24$ (solid circles), and the value of $\lambda$ at infinite size obtained by the BST extrapolation is indicated by an open circle with an error bar.}
%\label{figlam}
%\end{figure}
%
%
%
%
\end{document}